\documentclass[useAMS,usenatbib]{mn2e}
\usepackage{natbib}
\usepackage{amsmath,amsfonts}
\usepackage{epsfig}
\usepackage{mathptmx}
\usepackage{longtable}

\def\reff@jnl#1{{\rm#1\/}}

\def\aj{\reff@jnl{AJ}}                  
\def\araa{\reff@jnl{ARA\&A}}            
\def\apj{\reff@jnl{ApJ}}                        
\def\apjl{\reff@jnl{ApJ}}               
\def\apjs{\reff@jnl{ApJS}}              
\def\ao{\reff@jnl{Appl.Optics}}         
\def\apss{\reff@jnl{Ap\&SS}}            
\def\aap{\reff@jnl{A\&A}}                       
\def\apjl{\reff@jnl{ApJ}}               
\def\aapr{\reff@jnl{A\&A~Rev.}}         
\def\aaps{\reff@jnl{A\&AS}}             
\def\azh{\reff@jnl{AZh}}                        
\def\baas{\reff@jnl{BAAS}}              
\def\jrasc{\reff@jnl{JRASC}}            
\def\memras{\reff@jnl{MmRAS}}           
\def\mnras{\reff@jnl{MNRAS}}            
\def\pra{\reff@jnl{Phys. Rev. A}}         
\def\prb{\reff@jnl{Phys. Rev. B}}         
\def\prc{\reff@jnl{Phys. Rev. C}}         
\def\prd{\reff@jnl{Phys. Rev. D}}         
\def\prl{\reff@jnl{Phys. Rev. Lett}}      
\def\pasp{\reff@jnl{PASP}}              
\def\pasj{\reff@jnl{PASJ}}              
\def\qjras{\reff@jnl{QJRAS}}            
\def\skytel{\reff@jnl{S\&T}}            
\def\solphys{\reff@jnl{Solar~Phys.}}    
\def\sovast{\reff@jnl{Soviet~Ast.}}     
\def\ssr{\reff@jnl{Space~Sci.Rev.}}     
\def\zap{\reff@jnl{ZAp}}                        
\def\nat{\reff@jnl{Nature}}             

\def\p#1by#2{{\partial{#1} \over \partial{#2}}}
\def\pp#1by#2#3{{\partial^2{#1} \over \partial{#2}\partial{#3}}}
\def\d#1by#2{{{\rm d}{#1} \over {\rm d}{#2}}}
\def\dd#1by#2#3{{{\rm d}^2{#1} \over {\rm d}{#2}{\rm d}{#3}}}

\title[A new halo in Triangulum Australis]{KAT-7 detection of radio halo emission in the Triangulum Australis galaxy cluster}

\label{firstpage}

\author[Scaife et~al.]{
 Anna~M.~M.~Scaife$^1$\thanks{email: anna.scaife@manchester.ac.uk}, 
 Nadeem Oozeer$^{2,3,4}$, 
 Francesco de~Gasperin$^5$, 
 Marcus Br{\"u}ggen$^5$, 
\newauthor
 Cyril Tasse$^{2,6,7}$ 
 \& Lindsay Magnus$^2$.
 \vspace{0.03in}\\
$^1$ Jodrell Bank Centre for Astrophysics, University of Manchester, Manchester M13 9PL\\
$^2$ SKA South Africa, The Park, Park Road, Pinelands, 
     Cape Town 7405, South Africa\\
$^3$ African Institute for Mathematical Sciences, 
     6-8 Melrose Road, Muizenberg 7945, South Africa\\
$^4$ Centre for Space Research, North-West University, 
     Potchefstroom 2520, South Africa \\
$^5$ Hamburger Sternwarte, University of Hamburg, Gojenbergsweg 112, 21029 Hamburg, Germany\\
$^6$ GEPI, Observatoire de Paris, CNRS, Universit{\'e} Paris Diderot, 5 place Jules Janssen, 92190 Meudon, France\\
$^7$ Department of Physics \& Electronics, Rhodes University, PO Box 94, Grahamstown, 6140, South Africa\\
}

\date{Accepted ---; received ---; in original form \today}

\pagerange{\pageref{firstpage}--\pageref{lastpage}}

\pubyear{2010}

\begin{document}

\maketitle

\label{firstpage}

\begin{abstract}
We report the presence of high significance diffuse radio emission from the Triangulum Australis cluster using observations made with the KAT-7 telescope and propose that this emission is a giant radio halo. We compare the radio power from this proposed halo with X-ray and SZ measurements and demonstrate that it is consistent with the established scaling relations for cluster haloes. By combining the X-ray and SZ data we calculate the ratio of non-thermal to thermal electron pressure within Triangulum Australis to be $X=0.658\pm0.054$. We use this ratio to constrain the maximum magnetic field strength within the halo region to be $B_{\rm max, halo} = 33.08\,\mu$G and compare this with the minimum field strength from equipartition of $B_{\rm min, halo} = 0.77(1+k)^{2/7}\,\mu$G to place limits on the range of allowed magnetic field strength within this cluster. We compare these values to those for more well-studied systems and discuss these results in the context of equipartition of non-thermal energy densities within clusters of galaxies.

\end{abstract}


\section{Introduction}

A number of galaxy clusters are sources of diffuse radio emission that can be classified as either radio halos or radio relics (e.g., Feretti et al. 2012, and references therein). The radio emission is synchrotron radiation produced by relativistic electrons with Lorentz factors of the order of $10^4$ that move in $\mu$G magnetic fields.

Giant radio halos have sizes of $1 - 2$\,Mpc, are located at the centres of clusters, have fairly steep spectra and are not usually observed to have significant polarization (e.g., Feretti et al. 2001; Bacchi et al. 2003). Synchrotron emission from such large volumes requires that local particle acceleration is effective throughout the cluster (Jaffe 1977). Although the basic observational properties of radio halos have been established (e.g., Feretti et al. 2012), the formation mechanism of radio halos is still unclear (e.g., Brunetti et al. 2008; Donnert et al. 2010a,b; Macario et al. 2010; Brown \& Rudnick 2011; En{\ss}lin et al. 2011; Brunetti et al. 2012; Zandanel et al. 2012; Arlen et al. 2012). Theories that explain their origins include “primary” models, in which an existing electron population is re-accelerated by turbulence caused by recent cluster mergers (Brunetti et al. 2001; Petrosian 2001), and “secondary” models, in which relativistic electrons are continuously injected into the ICM by inelastic collisions between cosmic rays and thermal ions (e.g., Dennison 1980; Blasi \& Colafrancesco 1999; Dolag \& En{\ss}lin 2000; Miniati et al. 2001; Keshet \& Loeb 2010). Combinations of both acceleration mechanisms have also been considered (Brunetti \& Blasi 2005; Dolag et al. 2008; Brunetti \& Lazarian 2011).


\begin{figure*}
\centerline{\includegraphics[width=0.5\textwidth]{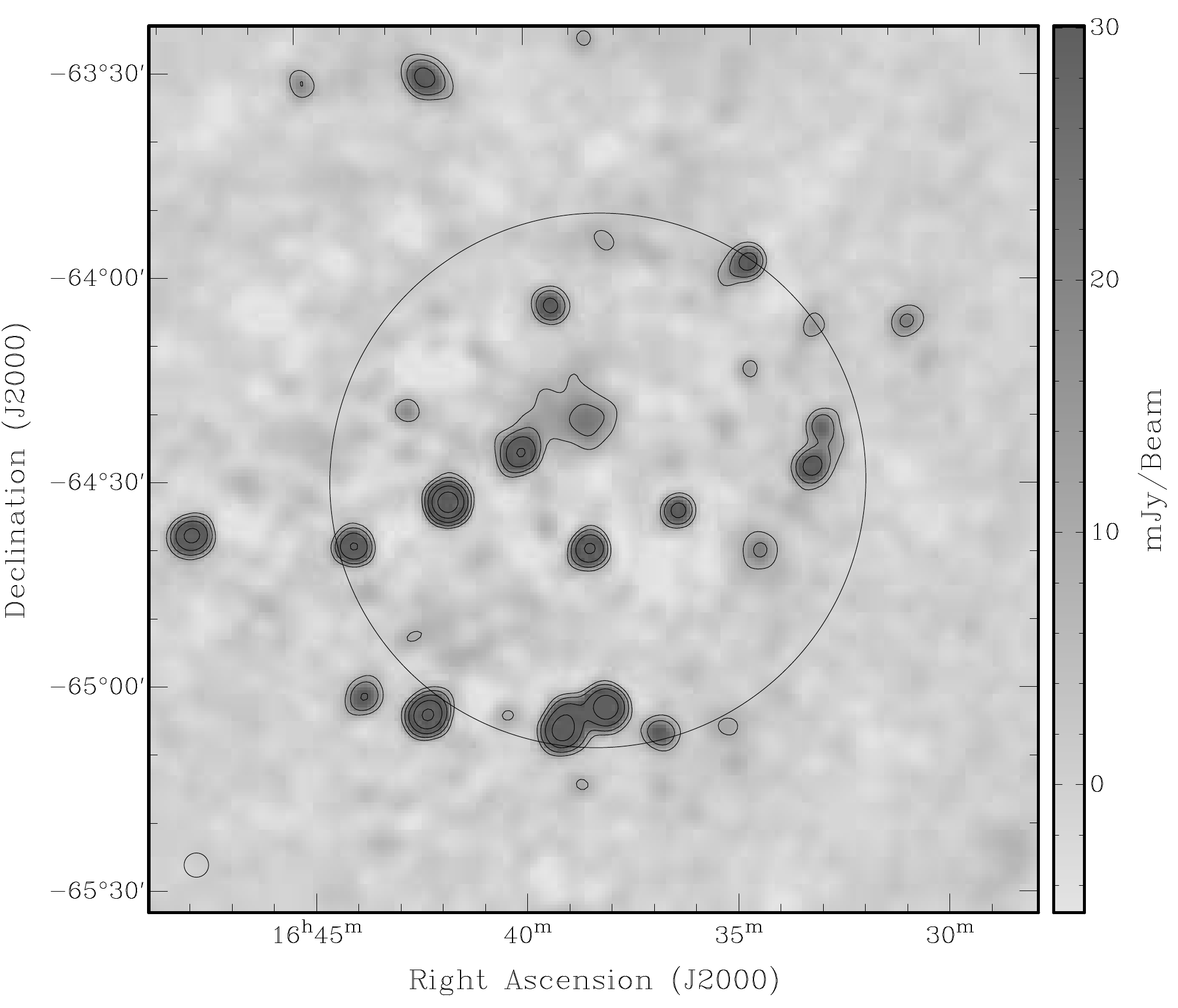}\qquad\includegraphics[width=0.485\textwidth]{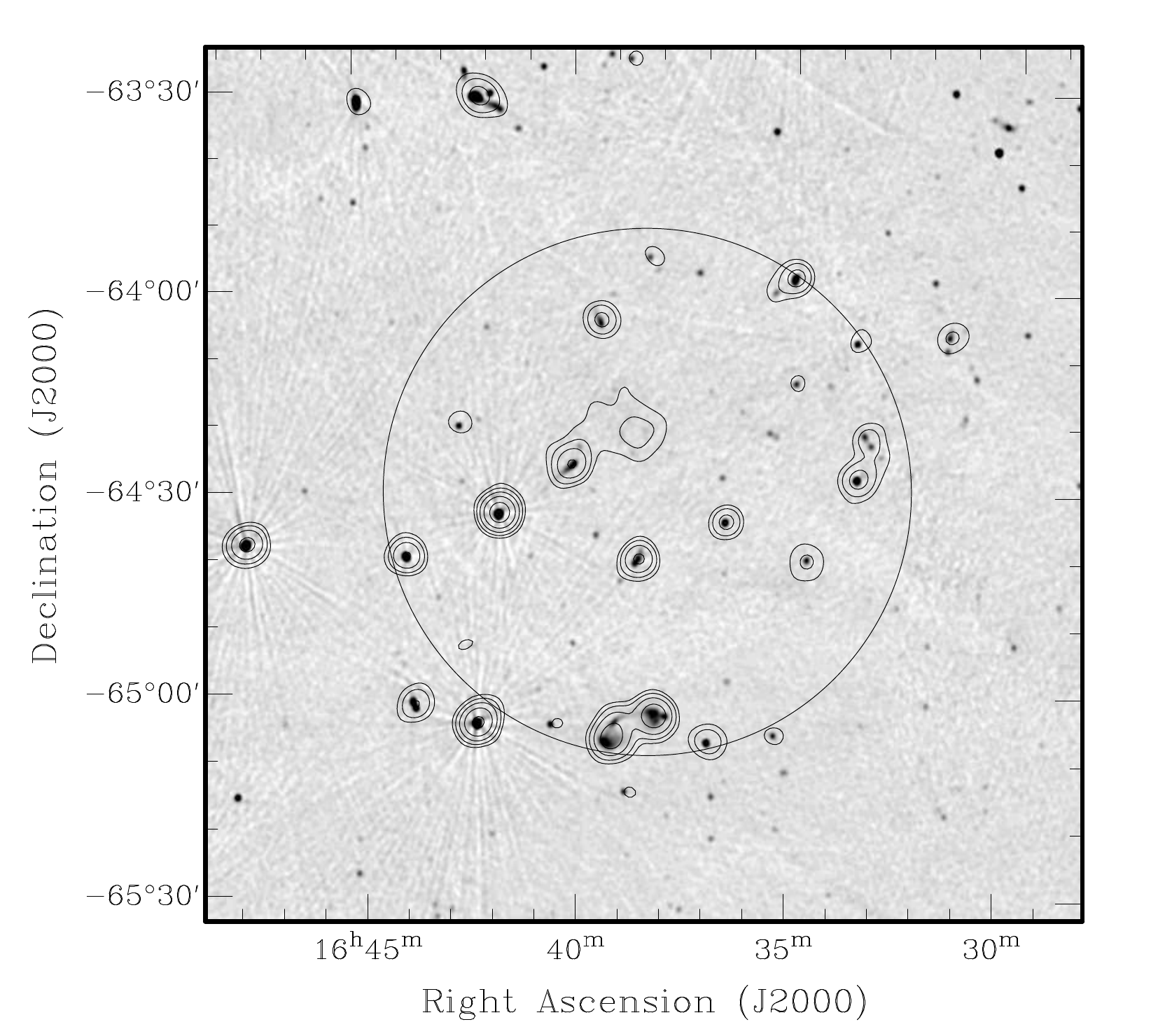}}
\caption{1328\,MHz KAT-7 image of the Triangulum Australis region. \emph{Left}: KAT-7 data are shown as greyscale and contours for the full field without point source subtraction. The half power point of the KAT-7 primary beam is shown as a circle and the synthesized beam is shown in the bottom left corner with dimensions of $3.67\times3.41$\,arcmin. Greyscale data are saturated at 30\,mJy\,beam$^{-1}$ in order to highlight the low surface brightness diffuse emission. Contours are shown in increments of 5\,$\sigma_{\rm rms}$ from 5\,$\sigma_{\rm rms}$. \emph{Right}: KAT-7 data are shown as contours, with intervals as in the left-hand figure; SUMSS data are shown as greyscale, saturated at 400\,mJy\,beam$^{-1}$ in order to highlight low surface brightness emission. The half power point of the KAT-7 primary beam is shown as a circle. In both maps the KAT-7 $\sigma_{\rm rms}=1.84$\,mJy\,beam$^{-1}$ and no correction is applied for the KAT-7 primary beam response.\label{fig:image}}
\end{figure*}

\begin{figure}
\centerline{\includegraphics[width=0.5\textwidth]{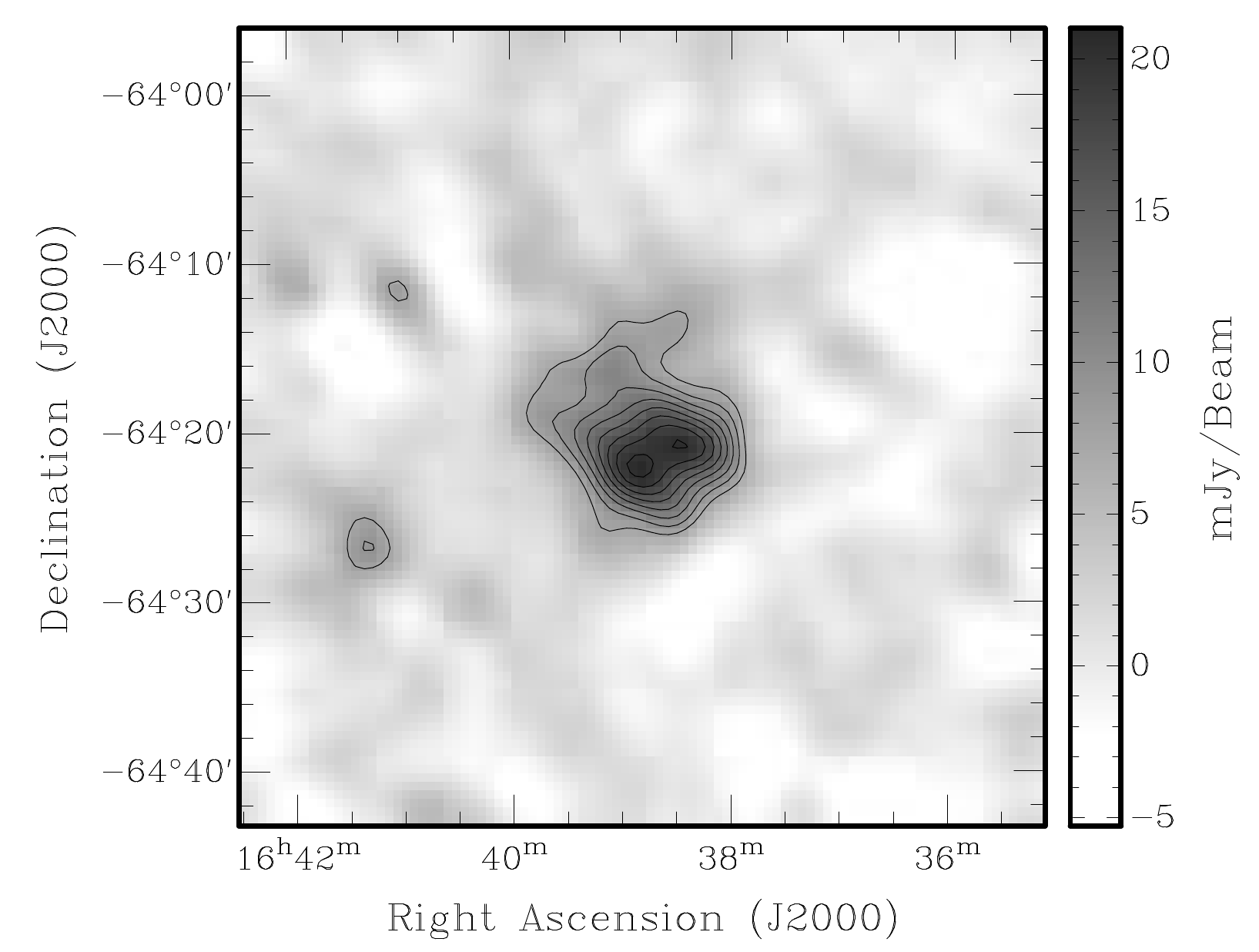}}
\caption{1328\,MHz point source subtracted KAT-7 image of the proposed halo emission within Triangulum Australis. Contours are shown in increments of 1\,$\sigma_{\rm rms}$ from 4\,$\sigma_{\rm rms}$, where $\sigma_{\rm rms}=1.84$\,mJy\,beam$^{-1}$. No correction for the primary beam response has been applied to these data.\label{fig:zoom}}
\end{figure}

Since few radio telescopes cover the very low declinations, most radio halos are found in the Northern sky. The only radio halo known below a declination of $-40$ deg is the bullet cluster (Liang et al. 2000.) In order to extend the sample of radio halos, we started from the BAX \footnote{http://bax.ast.obs-mip.fr} cluster catalog, selecting those objects with declination $<-40$ deg, $T>4$ keV, $z<0.5$, and some evidence of a merger either from the ROSAT images or from the literature. This resulted in a sample of eight clusters. In this paper we present the first of these: Triangulum Australis. The Triangulum Australis cluster is a relatively nearby ($z = 0.051$) bright, hot system, which was overlooked in the optical band due to its low Galactic latitude. It was first discovered as an X-ray source (McHardy et al. 1981). The Triangulum Australis cluster has been observed with XMM-Newton (60\,ks, Markevitch et al. 1996) and it was found that this cluster has a hot (12 keV) core at its centre that is most likely produced by a merger. 

Finally, the cluster is close enough ($z=0.051$, $5' \simeq 300$ kpc) that even at low resolution a radio halo could be resolved. As part of the development of MeerKAT (Booth et al. 2009), a scientific test array, the Karoo Array Telescope (KAT-7), has been constructed and commissioned at the same site. In this paper we report the discovery of a giant radio halo with the KAT-7 array, showing the potential of the array to image extended, low-surface brightness objects.

Throughout this paper we assume a $\Lambda$CDM cosmology with $H_{\circ}$ = 67.3\,km\,s$^{-1}$\,Mpc$^{-1}$, $\Omega_m$ = 0.32, and $\Omega_\Lambda$ = 0.68. All images are in the J2000 coordinate system and all errors are quoted at 1\,$\sigma$.

\section{Observations}\label{observations}

The KAT-7 telescope consists of seven 12\,m diameter dishes, equipped with cryogenically cooled receivers working between 1.3\,GHz and 1.8\,GHz with an observational bandwidth of 256\,MHz. The dish distribution is optimized for a Gaussian UV distribution, with highest weighting given to the optimisation parameters of 4-hour tracks at 60 degrees declination (de Villiers 2007). The maximum baseline separation is 192m and minimum spacing is 24m. 

Triangulum Australis was observed as part of general commissioning for the KAT-7 instrument four times between 2013 Feb and 2013 June at a central frequency of 1.328\,GHz, giving a total integration time of approximately 40\,hours. For each observation, primary calibration was performed using PKS\,1934-638, while secondary gain calibration used PKS\,1718-649. 

\subsection{Data reduction}

The native KAT-7 data comes in the Hierarchical Data Format (hdf5). Once converted into measurement set (ms) format using in-house software, the data were reduced using the CASA package\footnote{\tt www.nrao.edu/casapy}. Channels contaminated by known RFI were flagged immediately, followed by automated flagging using the CASA {\sc rflag} routine, looking at both auto- and cross-polarisation components. After flagging, the data were calibrated following standard practice. Flux densities were set using PKS\,1934-638, tied to the Perley-Butler-2010 flux density scale in {\sc setjy}.

MS-MFS deconvolution was carried out using the {\sc clean} task in CASA over a ~ 2$^\circ\;\times$ 2$^\circ$ field-of-view (FOV; 1.5 times the FWHM of the KAT-7 primary beam). Imaging was performed by initially using a mask based on sources from the SUMSS catalogue (Mauch et~al. 2003) with 843\,MHz flux densities exceeding 15\,mJy, before removing the mask to allow deconvolution of the whole field. The resulting Stokes~I image is shown in Fig.~\ref{fig:image} and has an rms noise of $\sigma_{\rm rms} = 1.84$\,mJy\,beam$^{-1}$, which is measured using the rms in the central region of the source subtracted image. These data are confusion limited at the resolution of KAT-7. Predictions of the expected confusion level for KAT-7 at this frequency are slightly lower than the measured rms noise in these data, $\sigma_{\rm conf} \simeq 1.4$\,mJy\,beam$^{-1}$ (Riseley et~al. 2014). We attribute this difference to the enhanced source population towards galaxy clusters, relative to the field.



\section{Results}

Diffuse emission towards the Triangulum Australis cluster is visible in the KAT-7 data at a significance of $>10\,\sigma$ over an extent of several arcminutes and a major axis of approximately 1\,Mpc within the 5\,$\sigma$ contour, see Fig.~\ref{fig:image}. The diffuse radio emission is coincident with the X-ray emission towards this cluster, although a slight offset ($\approx 2$\,arcmin) exists between the peaks of the radio and X-ray emission. The KAT-7 image has astrometry for the radio point source population in this field consistent with previous high resolution surveys (SUMSS; Mauch et~al. 2003), see Fig.~\ref{fig:image}\,(right), which suggests that this offset is due to the differing nature of the physical processes responsible for the X-ray and radio emission: X-ray emission predominantly traces the density of the thermal gas population within the cluster, whereas radio emission traces the cosmic ray electron population and magnetic field strength distribution. Offsets between the peak surface brightness of different emission mechanisms in disturbed clusters are not uncommon, see e.g. Rodr{\'i}guez-Gonz{\'a}lvez et~al. (2011). We do not make a further physical interpretation of this offset in this work. 

\begin{table*}
\caption{Sources detected at a significance of $\geq 7.5\,\sigma_{\rm rms}$ within the half power point of the KAT-7 primary beam with SUMSS counterparts. Column [1] lists a numerical designation for each source; column [2] lists the fitted Right Ascension, with the fit error on this position listed in column [3]; column [4] lists the fitted Declination, with the fit error on this position listed in column [5]; column [6] lists the fitted peak flux density for each source from the KAT-7 data; where sources are unresolved this value is listed as ``-"; column [7] lists the integrated flux density for each source from the KAT-7 data; column [8] lists the SUMSS flux density for coincident sources; column [9] lists the SUMSS source designation for the listed SUMSS flux densities where names are composed of the truncated SUMSS J2000 coordinates; column [10] lists the measured spectral index between the SUMSS and KAT-7 flux densities; column [11] identifies notes on the KAT-7 data fitting. \label{tab:sources}}
\begin{tabular}{lccccclllll}
\hline\hline
No. & RA && Dec && $S_{\rm peak, 1328}$ & $S_{\rm int, 1328}$ & $S_{\rm int, 843}$ & SUMSS & $\alpha_{843}^{1328}$ & Notes \\
 & [J2000] & [s] & [J2000] & [arcsec] & [mJy\,bm$^{-1}$] & [mJy] & [mJy] & Identifier & & \\
\hline
 001 & 16 33 13.2 & $\pm0.31$ & $-$64 23 00 & $\pm4.69$ & 47.4$\pm$3.8  & 82.9$\pm$6.2  & \hskip -0.12in $\left \{\begin{tabular}{ll} 34.7$\pm$1.5\\$50.9\pm3.1$ \end{tabular} \right. $ & \hskip -0.11in $\left. \begin{tabular}{ll} J163309-642322 \\ J163317-642157 \end{tabular} \right \}$ &$0.07\pm0.19$  & (1;2) \\
 &&&&&&&&&&\\
 002 & 16 33 25.8 & $\pm0.42$ & $-$64 28 22 & $\pm4.33$ & 77.5$\pm$5.2 & 116.5$\pm$7.6 & 131.6$\pm$4.1 & J163327-642832 & $0.27\pm0.16$&  (2) \\
 003 & 16 33 28.7 & $\pm0.54$ & $-$64 07 46 & $\pm8.19$ & 27.8$\pm$2.1  & 38.4$\pm$2.9 & 60.8$\pm$2.1 & J163331-640805 & $1.01\pm0.18$&  (2) \\
 004 & 16 34 37.7 & $\pm0.38$ & $-$64 41 12 & $\pm5.72$ & 27.6$\pm$2.5  & 54.9$\pm$3.7  & 47.8$\pm$1.8 & J163434-644040 & $-0.30\pm0.17$ & (2) \\
 005 & 16 34 53.4 & $\pm0.52$ & $-$64 14 25 & $\pm7.79$ & $-$  & 19.1$\pm$2.5  & 40.8$\pm$1.6 &J163453-641420 &  $1.67\pm0.30$ & \\
 &&&&&&&&&&\\
 006 & 16 35 01.1 & $\pm0.42$ & $-$63 58 44 & $\pm6.32$ & 98.5$\pm$5.8  & 117.7$\pm$6.9  & \hskip -0.12in $\left \{\begin{tabular}{ll} 152.4$\pm$6.0\\$29.9\pm2.9$ \end{tabular} \right. $ & \hskip -0.11in $\left. \begin{tabular}{ll} J163457-635838 \\ J163523-640040 \end{tabular} \right \}$ &$0.96\pm0.15$  & (2) \\
 &&&&&&&&&&\\
 007 & 16 36 30.0 & $\pm0.20$ & $-$64 35 22 & $\pm2.96$ & $-$  & 54.1$\pm$3.6 & 90.8$\pm$2.9 & J163629-643515 & $1.14\pm0.16$ & \\
 008 & 16 36 51.7 & $\pm0.25$ & $-$65 08 07 & $\pm3.75$ & 66.1$\pm$4.4  & 82.4$\pm$5.4  & 101.3$\pm$3.2 & J163652-650808  &    $0.45\pm0.16$         & \\ 
 &&&&&&&&&&\\
 009 & 16 38 10.5 & $\pm0.44$ & $-$65 04 29 & $\pm6.64$ & 198.0$\pm$13.3 & 290.0$\pm$19.6  & \hskip -0.12in $\left \{\begin{tabular}{ll} 347.9$\pm$13.9\\47.0$\pm$1.9 \end{tabular} \right. $ & \hskip -0.11in $\left. \begin{tabular}{ll} J163808-650409\\ J163751-650414 \end{tabular}\right \}$ & $0.68\pm0.18$ & (3) \\
 &&&&&&&&&&\\
 010 & 16 38 13.1 & $\pm0.36$ & $-$63 55 19 & $\pm5.44$ & 16.6$\pm$1.0  & 27.5$\pm$2.3 & 32.6$\pm$1.4 & J163816-635536 & $0.37\pm0.21$&  (3) \\
 011 & 16 38 31.8 & $\pm0.19$ & $-$64 41 04 & $\pm$2.90 & 89.0$\pm$5.1  & 92.6$\pm$5.3  & 105.0$\pm$5.5 &J163830-644043 & $0.28\pm0.17$ & \\
 &&&&&&&&&&\\
 012 & 16 39 07.5 & $\pm0.25$ & $-$65 07 20 & $\pm3.79$ & 203.8$\pm$12.6 & 337.0$\pm$20.8  & \hskip -0.12in $\left \{ \begin{tabular}{ll} 339.6$\pm$13.6\\95.1$\pm$6.1 \end{tabular} \right. $ & \hskip -0.11in $\left. \begin{tabular}{ll} J163913-650804\\J163903-650513 \end{tabular}\right \}$ & $0.56\pm0.16$ & (3) \\
 &&&&&&&&&&\\
 013 & 16 39 24.2 & $\pm0.10$ & $-$64 05 13 & $\pm1.50$ & 64.7$\pm$3.8  & 68.1$\pm$4.0  & 89.7$\pm$4.5 &J163924-640520 & $0.61\pm0.17$ & \\
 &&&&&&&&&&\\
 014 & 16 40 05.5 & $\pm0.23$ & $-$64 26 42 & $\pm$3.43 & 79.2$\pm$5.9  & 124.5$\pm$8.9 & \hskip -0.12in $\left \{ \begin{tabular}{ll} 47.6$\pm$3.0 \\96.4$\pm$5.7 \end{tabular} \right. $ & \hskip -0.11in $\left. \begin{tabular}{ll} J164000-642639 \\J164007-642717\end{tabular}\right \}$ & $0.32\pm0.19$ & (3;4)\\
 &&&&&&&&&&\\
 015 & 16 40 24.3 & $\pm0.96$ & $-$65 05 35 & $\pm$14.43& $-$  & 32.2$\pm$3.0  & 55.9$\pm$2.1  &J164033-650529 & $1.21\pm0.22$ & \\
 016 & 16 41 45.9 & $\pm0.05$ & $-$64 34 02 & $\pm0.77$ & 296.2$\pm$15.1 & 302.4$\pm$15.4 & 443.5$\pm$13.4 &J164145-643407 & $0.84\pm0.13$ &\\
 017 & 16 42 38.4 & $\pm0.36$ & $-$64 20 30 & $\pm5.38$ & $-$ & 26.9$\pm$2.6   & 53.6$\pm$1.9 & J164239-642050 & $1.51\pm0.23$ & \\
 018 & 16 43 55.8 & $\pm0.11$ & $-$64 40 16 & $\pm1.63$ & 148.8$\pm$8.2  & 154.8$\pm$8.5  & 238.9$\pm$7.3 &J164354-644019 & $0.95\pm0.14$ &\\
\hline
\end{tabular}
\begin{minipage}{\textwidth}
\noindent
Note 1: Adjacent source SUMSS\,J163254-643254 may also contribute emission at 843\,MHz. Fitting may be affected.

\noindent
Note 2: Diffuse component evident in KAT-7 data.

\noindent
Note 3: Closely adjacent source. Fitting may be affected.

\noindent
Note 4: Additional component listed in original SUMSS catalogue (Mauch et~al. 2003) but not in later version (Mauch et~al. 2008).

\end{minipage}
\end{table*}

From X-ray studies, it has previously been proposed that Triangulum Australis is a merging system due to its high central temperature (Markevitch et~al. 1996) and therefore likely to host a giant radio halo. We propose here that the diffuse radio emission detected with KAT-7 is associated with that halo. It is possible that the extension of the radio emission seen towards the North of the cluster may be due in part to an unresolved cluster relic; however, given the low significance of this protrusion we do not try to separate these features. We note that neither diffuse nor compact emission is present in SUMSS (Mauch et~al. 2003) data towards the proposed halo emission. This reduces the possibility that the emission detected with KAT-7 is due to a collection of unresolved point sources.

\subsection{Point source removal}

A large number of point sources are also detected within the KAT-7 field of view, see Fig.~\ref{fig:image}. In order to reduce any confusing effect on the diffuse emission identified with the halo, these sources were used to solve for direction dependent calibration solutions before being subtracted from the visibility dataset. At the frequencies of our observations, the cross-correlation between
voltages from pairs of antenna are affected by a series of moderate
but complex baseline-time-frequency Direction Dependent Effects
(DDE). They might include atmospheric effects, pointing errors or dish
deformation.

A large variety of solvers have been developed to tackle these kinds of calibration issues. Here we do not attempt to physically characterize the DDEs but instead use a Jones-based solver. This type of solver constitutes the most widely used family of algorithms for direction dependent calibration, and aims at estimating the {\it apparent} net product of the various effects mentioned above. Recently, algorithms have been developed (see e.g. Yatawatta et~al. 2008; Noordam et~al. 2010), that estimate a Jones matrix per time-frequency bin per antenna, per direction. The well known problems associated with this type of technique are (i) ill-conditioning and (ii) computational cost, both being due to the larger number of degrees of freedom used to solve the problem (compared to the direction-independent case). The first of these issues can affect the scientific signal by suppressing un-modelled flux, while the cubic computational cost with the number of degrees of freedom can put strong limitations on the affordable number of direction-dependent parameters.

The Jones-based solver utilized here (Tasse et~al. 2014) is a DDE variant of the StefCal approach (Salvini et~al. 2014). It operates using the concept of iteratively solving for linear systems in a similar manner to traditional non-linear least-squares solvers, but by using an alternative iteration scheme, significantly improving convergence speed and robustness.

For the data presented here, in order to increase the signal in each direction, we clustered the sources in 5 direction-based groups by using a Voronoi tessellation algorithm and computed a scalar direction dependent Jones matrix every 15 minutes. We verified that this strategy was not driving suppression of the unmodelled flux by using incomplete sky models. 

Following direction-dependent calibration, point sources above a significance of 7.5\,$\sigma_{\rm rms}$ were then subtracted directly from the visibility data. Point source subtracted data were then re-imaged using natural weighting in order to enhance the signal-to-noise of the low surface brightness halo on large scales. The point-source-subtracted image is shown in Fig.~\ref{fig:zoom}. 

Sources with flux densities above 7.5\,$\sigma_{\rm rms}$ within the 50\,per~cent power point of the KAT-7 primary beam are listed in Table~\ref{tab:sources}, where they are cross-referenced with the v2.1 SUMSS catalogue (Mauch et~al. 2008; hereafter M08). Errors on KAT-7 flux densities as listed in Table~\ref{tab:sources} are calculated as $\sigma = \sqrt{\sigma_{\rm rms}^2 + \sigma_{\rm fit}^2 + (0.05S_{\rm fit})^2}$. Where multiple SUMSS sources are associated with a single KAT-7 detection, due to the large difference in resolution between these two surveys, the combined flux density is used to calculate the spectral index value; in this case, uncertainties on the SUMSS data points are combined in quadrature. The average spectral index for the sources in this list is $\bar{\alpha}_{\rm 843\,MHz}^{\rm 1328\,MHz} = 0.66\pm0.43$, typical of optically thin non-thermal emission. 

\subsection{Flux density estimation}

Integrated flux density measurements for the proposed halo were made using the source subtracted images in order to avoid contamination from the point source population. The region of diffuse emission that we associate here with the radio halo of Triangulum Australis is extended and irregular. The centroid of the diffuse emission is located at J\,16h38m48.5s~$-$64$^{\circ}$20$'$13$''$ and the peak of the diffuse radio emission at J\,16h38m52.5s~$-$64$^{\circ}$22$'$01$''$, see Figure~\ref{fig:image}. Within the 3\,$\sigma_{\rm rms}$ contour the diffuse emission has dimensions of $850\times990$\,arcsec (East-West by North-South) and a major axis of 1022\,arcsec, where 1\,arcsec is 1.035\,kpc at $z=0.051$.  The CASA task {\sc imfit} applied to this target returns a value for the integrated flux density of $S_{\rm imfit} = 186\pm15$\,mJy. However, this method involves fitting a Gaussian to our target region, which is significantly non-Gaussian in morphology. Consequently we also use an aperture photometry technique to extract the integrated flux density using the {\sc fitflux} code (Green 2007). This method fits a tilted plane to the edges of a defined aperture before subtracting this plane and integrating the remaining flux density. Using this method on data corrected for a Gaussian primary beam with a FWHM of 1.31\,degrees, we find an integrated flux density for the radio halo of $S_{\rm fitflux} = 130\pm4$\,mJy, where the error on the fitted value is calculated using the standard deviation of the recovered flux density from multiple apertures of varying dimension. We calculate our complete uncertainty on the integrated flux density as $\sigma = \sqrt{\sigma_{\rm rms}^2 + \sigma_{\rm fitflux}^2 + (0.05S_{\rm int, fitflux})^2}$ to give a final integrated flux density of $S_{\rm int, halo} = 130\pm8$\,mJy.

\section{Scaling Relations}
\label{sec:scaling}

The power-law relationship between radio power, $P_{\rm rad}$, and X-ray luminosity, $L_{\rm X}$, is well known for clusters hosting haloes (e.g. Giovannini et~al. 2000; Feretti et~al. 2012) and is commonly characterised as $P_{\rm rad} \propto L_{\rm X}^d$, where $d$ has values of approximately $1.5-2.1$ (Brunetti et~al. 2009). A further power-law relationship is also known linking radio power and the integrated Compton-$y$ parameter, $Y_{\rm SZ}$, determined from observations of the Sunyaev-Zel'dovich (SZ) effect. Unlike the X-ray luminosity, which depends on the propertes of the thermal components within the cluster, $Y_{\rm SZ}$ is proportional to the total electron pressure integrated along the line of sight (Colafrancesco et~al. 2003). Consequently, the correlation of radio power to integrated Compton-$y$ is of particular interest as it indicates the relationship between the non-thermal electron pressure component (characterised by $P_{\rm rad} \propto n_{\rm e,rel}B^{(\alpha+1)/2}\nu^{-(\alpha+1)/2} \sim P_{\rm non-thermal} U_{\rm B}^{(\alpha+1)/4}$) and the total electron pressure. Following Colafrancesco et~al. (2014), we denote the ratio of these quantities as $X$, where
\begin{equation}
P_{\rm total} = P_{\rm thermal} + P_{\rm non-thermal} = (1+X)P_{\rm thermal}.
\end{equation}

\begin{figure}
\centerline{\includegraphics[width=0.5\textwidth]{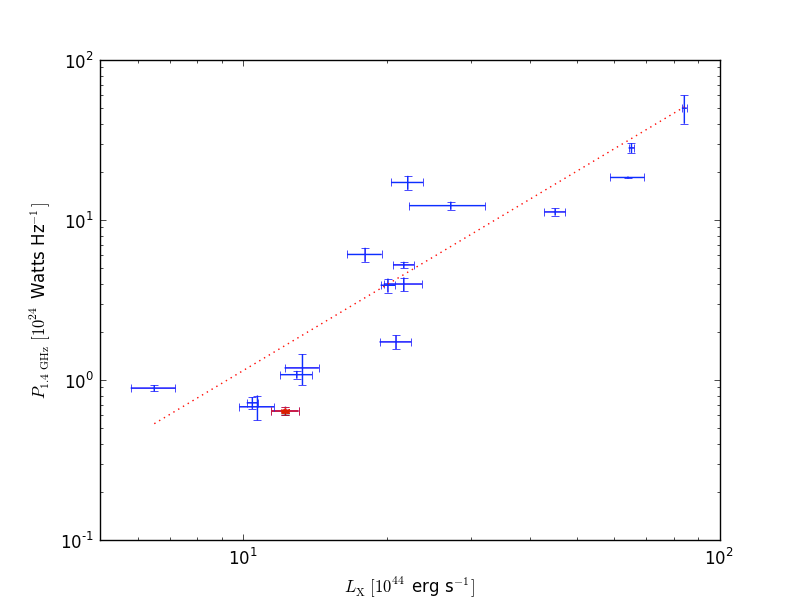}}
\centerline{\includegraphics[width=0.5\textwidth]{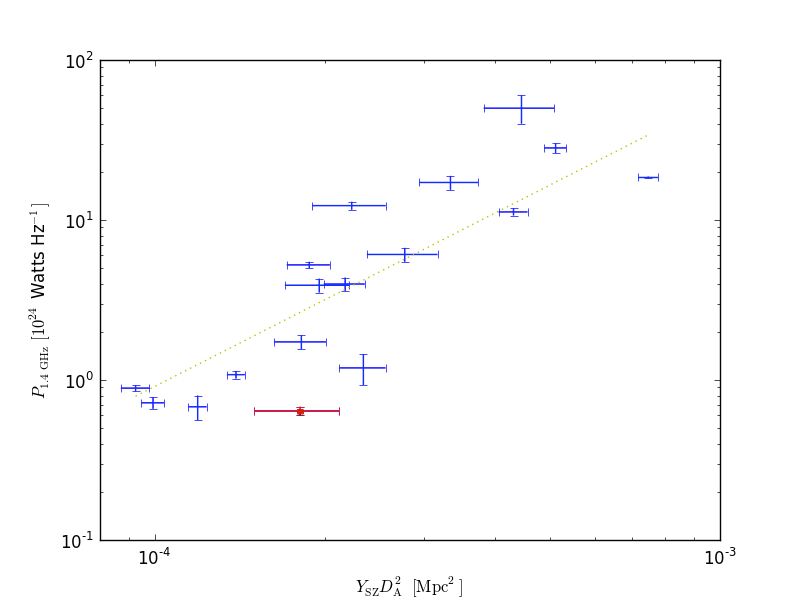}}
\caption{Cluster halo scaling relations. Top: X-ray luminosity and radio power scaling relation; Bottom: Integrated Compton-$y$ and radio power scaling relation. Data are taken from Colafrancesco et~al. (2014; black points) with the exception of Triangulum Australis (red squares), which has properties as determined in this work. The fitted power-law relations are taken from Colafrancesco et~al. (2014) and shown as dashed lines.}\label{fig:scaling}
\end{figure}

As can be seen from Figure~\ref{fig:scaling}, the radio power from Triangulum Australis is consistent with the known scaling relations. Here we use the sample of Colafrancesco et~al. (2014), which extends the sample of Basu et~al. (2009). Integrated Compton-$y$ values are taken from the Planck catalogue (Planck Collaboration 2013), which measures the cylindrical volume integrated Compton-$y$, $Y_{\rm cyl} ®= Y_{\rm SZ}D_{\rm A}^2$, within an aperture of $R=5R_{500}$. At this radius the cylindrical integrated quantity is equivalent to the spherically integrated quantity, $Y_{\rm sph}$ (Arnaud et~al. 2010). Furthermore, it can then be trivially related to $Y_{\rm R500}$ as $Y_{\rm R500} = Y_{\rm 5R500} \times I(1)/I(5)$, where $I(1)=0.6552$ and $I(5)=1.1885$ (Appendix 2 of Arnaud et~al. 2010).

\subsection{Non-thermal pressure fraction}

The value of the ratio $X$ can be determined from the X-ray luminosity, $L_{\rm X}$ and the integrated Compton-$y$ value,  as calculated at $R_{500}$, the radius at which the cluster density profile is equal to 500 times the critical density of the Universe, $\rho_{\rm crit}(z) = 3H^2(z)/8\pi G$, where $H(z)=H_0E(z)$ and $E(z)=[\Omega_m (1+z)^3+\Omega_{\Lambda}]^{1/2}$. Again following Colafrancesco et~al. (2014), this relationship is,
\begin{equation}
Y_{\rm sph, R500} E(z)^{9/4} = \left[ \frac{(1+X)Y_0}{L_0^{5/4}}\right]L_X^{5/4},
\end{equation}
where the constants $Y_0$ and $L_0$ may be found from
\begin{eqnarray}
Y_0 &=& \frac{8\pi^2}{3}\frac{\sigma_{\rm T}}{m_{\rm e}c^2}G\mu m_{\rm p} 500 \rho_{\rm crit} n_{\rm e0,g} V_1(\lambda) \\
L_0 &=& 4\pi C_2 \left(\frac{2\pi}{3k_{\rm B}}G\mu m_{\rm p} 500 \rho_{\rm crit}\right)^{1/2}n^2_{\rm e0,g}\lambda^3 W_1(\lambda),
\end{eqnarray}
where $\mu=1.14$ is the mean molecular weight, $G$ is the gravitational constant, $m_{\rm p}$ is the proton mass, $\sigma_{\rm T}$ is the Thomson scattering cross-section and $C_2$ has the value $1.728\times 10^{40}$\,W\,s$^{-1}$\,K$^{-1/2}$\,m$^3$ (Rybicki \& Lightman 1985) with
\begin{eqnarray}
V_1(\lambda) &=& \int_0^{1/\lambda}{\left(1+u^2\right)^{-3\beta/2}u^2{\rm d}u},\\
W_1(\lambda) &=& \int_0^{1/\lambda}{\left(1+u^2\right)^{-3\beta}u^2{\rm d}u}.
\end{eqnarray}
This calculation assumes that the global cluster density profile is modelled by a $\beta$-model, with index $\beta$ and $r_{\rm c}=\lambda R_{500}$ and therefore has a central electron number density of
\begin{equation}
n_{\rm e0,g} = \frac{3\beta f_{\rm B}500\rho_{\rm crit}}{2\lambda^2 \mu_{\rm e} m_{\rm p}},
\end{equation}
where $f_{\rm B}$ is the baryon fraction, here assumed to have the value $f_{\rm B}=0.175$ (Planck Collaboration 2013). 

For Triangulum Australis we use $\beta=0.63\pm0.02$, $r_{\rm c}=3.5\pm0.2$\,arcmin  (Markevitch et~al. 1996) and $\lambda=0.3$, consistent with Colafrancesco et~al. (2014). Combining these with the X-ray luminosity and integrated Compton-$y$, we find that $X=0.658\pm0.054$.


\subsection{Maximum magnetic field strength}

The thermal electron pressure of the cluster within $R_{\rm 500}$ is expressed as
\begin{eqnarray}
\label{equ:bfromt} P_{\rm th, 500} &=& n_{\rm e, 500}k_{\rm B} T_{\rm 500}\\
\label{equ:bfromy}	    &=& \frac{m_{\rm e}c^2}{\sigma_{\rm T}}\frac{3}{4\pi}Y_{\rm sph, R500}R_{\rm 500}^{-3}\left( 1+ X \right)^{-1}.
\end{eqnarray}
Consequently, the average non-thermal pressure within $R_{\rm 500}$ can be calculated as $P_{\rm non-th, 500} = X P_{\rm th, 500}$. This additional pressure contribution to the SZ effect will come from the non-thermal particle population, with other kinetic forms of non-thermal pressure such as tubulence and bulk motions contributing to the kinetic SZ (kSZ) effect. Turbulence is generally assumed to be the dominant form of non-thermal pressure (Vazza et~al. 2012), but due to the directional nature of the kSZ effect and the multiple line of sight reversals expected for a turbulent medium, the net turbulent contribution to the kSZ effect is expected to be negligible, as is that of bulk motions (Sunyaev et~al. 2003). 

Furthermore, it is expected that magnetic pressure is sub-dominant and will not be greater than non-thermal particle pressure (e.g. Lagan{\'a} et~al. 2010; Brunetti \& Jones 2014). Under this assumption, one may calculate an upper limit on the strength of the cluster magnetic field such that
\begin{equation}
\label{eq:bcalc}
B_{\rm max, 500} \leq \sqrt{{8\pi}P_{\rm non-th, 500}},
\end{equation}
where $B$ is the magnetic field strength in Gauss and pressure is measured in Barye. From Eq.~\ref{equ:bfromt}, using a representative temperature of $T_{500} = 10$\,keV (Markevitch et~al. 1996) and noting that $n_{\rm e, 500}=500f_{\rm B}\rho_{\rm c}(z)/\mu_{\rm e}m_{\rm p}$, this provides an upper limit on the average magnetic field strength of $\langle B\rangle_{\rm max, 500} = 14.50\,\mu$G. Alternatively, using the integrated Compton-$Y$ value and Eq.~\ref{equ:bfromy}, this provides an upper limit on the average magnetic field strength within $R_{\rm 500}$ for Triangulum Australis of $\langle B\rangle_{\rm max, 500} = 19.46\,\mu$G.

Since the magnetic field strength is expected to vary as a function of cluster radius we convert our value of $\langle B\rangle_{\rm max, 500}$ to be more representative of the field within the halo region, which has a radius of 5\,arcmin, less than 50\% $R_{\rm 500}$. In order to do this we assume that the dependence of magnetic field strength on density follows
\begin{equation}
B(r) \propto \left(\frac{n_{\rm e}(r)}{n_{\rm e,0}}\right)^{\eta},
\end{equation}
(Bonafede et~al. 2013) where $\eta=0.5$, consistent with that determined for the Coma cluster. Combining this dependence with the standard $\beta$-model formalism for the radial density distribution, this gives
\begin{equation}
\label{eq:br}
B(r) \propto \left( 1+(r/r_{\rm c})^2\right)^{-3\beta/4}.
\end{equation}
The magnetic field strength determined from Eq.~\ref{eq:bcalc} is a volume-averaged quantity within $R_{\rm 500}$. Assuming spherical symmetry and $\langle B\rangle_{\rm max, 500} = 19.46\,\mu$G, the maximum average magnetic field strength within the halo radius, $r_{\rm h}$, is given by 
%
%
$\langle B\rangle_{\rm max, halo}=33.08\,\mu$G.


\subsection{Minimum magnetic field strength}

The minimum magnetic field strength, $B_{\rm min}$ in Tesla, within the halo region can be calculated, assuming equipartition, from its radio power following
\begin{equation}
B_{\rm min} = \left[ \frac{3\mu_0}{2}\frac{G(\alpha)(1+k)P_{\rm rad}}{Vf}\right]^{2/7}
\end{equation}
(Longair 2011) where $k$ is the ratio between the energy of heavy particles (protons) and the electrons, $f$ is the filling factor used to describe the fraction of the volume, $V$, occupied by radio emitting material and $G(\alpha)$ is defined as 
\begin{eqnarray}
\nonumber G(\alpha) &=& \frac{1}{a(p)(p-2)}\left[ \nu_{\rm min}^{-(p-2)/2} - \nu_{\rm max}^{-(p-2)/2}\right] \nu^{(p-1)/2}\\
\nonumber &&\times \frac{(7.4126\times 10^{-19})^{-(p-2)}}{2.344\times 10^{-25}}(1.253\times 10^{37})^{-(p-1)/2}\\
&&
\end{eqnarray}
with
\begin{equation}
a(p) = \frac{\sqrt{\pi}}{2}\frac{ \Gamma(p/4+19/12)\Gamma(p/4-1/12)\Gamma(p/4+5/4)}{(p+1)\Gamma(p/4+7/4)},
\end{equation}
where $p=1-2\alpha$. Since we only have a measurement of the radio power at a single frequency, we must assume a spectral index, $\alpha$. Here we use $\alpha=-1.5$, consistent with previously measured halo indices (Feretti et~al. 2012). 
This gives $p=4.0$ and $a(p)=0.034$; using $\nu_{\rm min}=10$\,MHz and  $\nu_{\rm max}=100$\,GHz we find $G(\alpha) = 0.256\nu^{1.5}$. For our measured flux density of $S_{\rm 1.33\,GHz} = 130\pm8$\,mJy, given the redshift of $z=0.051$, the radio power is $P_{\rm rad} = 0.6\times 10^{24}$\,W\,Hz$^{-1}$. For clusters of galaxies a value of $k=0$ or $k=1$ is typically used (Beck \& Krause 2005); however, the same authors also propose that in fact a larger value of $k$ ($k>>1$; such that $n_{\rm p}/n_{\rm e} \simeq 100-300$) is preferred by current models of cosmic ray production in galaxy clusters. Recent constraints using a combination  of radio data and upper limits from gamma-ray observations has shown that in galaxy clusters, $n_{\rm p}/n_{\rm e}$ is likely to be significantly less than 100 (Vazza \& Br{\"u}ggen 2014; see also  Guo, Sironi \& Narayan 2014). Here we model the radio halo as a solid sphere with radius, $r_{\rm h} = 5$\,arcmin. We assume that the volume is filled uniformly and completely, such that $f=1$. We find that, given these assumptions, $B_{\rm min} = 0.77(1+k)^{2/7}\,\mu$G.

We note that although equipartition and minimum energy arguments are frequently used, they are subject to a number of strong assumptions. One of the strongest assumptions is the value of the parameter $k$, the effect of which we have explicitly factored in our estimates. A further issue is the strong local dependence of the radio emissivity on magnetic field strength, which can cause $B_{\rm min}$ to over-estimate the volume-average field strength for inhomogeneous magnetic fields. A more complete discussion of these assumptions is presented in Beck \& Krause (2005).

\section{Discussion}

From considering the combination of X-ray, SZ and radio data we are able to place both lower and upper limits on the magnetic field strength in the halo region. The upper limit in this case assumes that the magnetic pressure will not be greater than all non-thermal particle pressures. Our derived upper limit for the magnetic field in the halo region of Triangulum Australis is $\langle B\rangle_{\rm max, halo}=33.08\,\mu$G. Since the cooling time (via synchrotron emission) of electrons is a function of magnetic field strength,
\begin{equation}
t_{\rm cool} = 0.23\left(\frac{\nu}{\rm 1.4\,GHz}\right)^{-0.5}\left(\frac{B}{B_{\rm CMB}}\right)^{-1.5}~{\rm Gyr},
\end{equation}
where $B_{\rm CMB} \simeq 3(1+z)^2\,\mu$G is the energy density equivalent magnetic field strength of the CMB, this field strength would imply a short synchrotron cooling time of $\simeq$7\,Myr. If the relativistic electrons of the radio halo are re-accelerated by first-order Fermi processes (as in Brunetti et al. 2001), the short loss time implies that the electron distribution evolves quickly, leading to an expedient decrease of the break energy of the electron spectrum. The short synchrotron loss time indicates that whatever mechanism powers the radio halo must still be active. For hadronic models, a short cooling time means that the radio emission must follow the ICM density distribution quite closely. This could potentially be probed by radio observations at higher angular resolution. For models of turbulent re-acceleration, it implies that the turbulence should have a high filling factor and be efficient in electron re-acceleration. Here, low-frequency observations would be useful (see e.g. Figs.~7 and~9 in Brunetti et al. 2001).

For comparison we consider the Coma cluster of galaxies (A1656), which is a particularly well studied system with extensive ancillary data available. Using $\beta$-model parameters from Briel, Henry \& B{\"o}hringer (1992), such that $\beta=0.75\pm0.03$ and $r_{\rm c}=10.5\pm0.6$\,arcmin, an X-ray luminosity of $(10.44\pm0.28)\times 10^{44}$\,erg\,s$^{-1}$ (Reichert et~al. 2011), an integrated Compton-$y$ of $(0.1173\pm0.0054)$\,arcmin$^2$ (Planck Collaboration 2011), a 1.4\,GHz radio halo power of $(0.72\pm0.06)\times 10^{24}$\,W\,Hz$^{-1}$ (Brunetti 2009), a halo radius of $r_{\rm h}=21$\,arcmin (Venturi et~al. 1990) and assumptions consistent with those outlined above we find $B_{\rm min, eq} = 0.46(1+k)^{2/7}\,\mu$G, $\langle B\rangle_{\rm max, R500} = 10.81\,\mu$G and $\langle B\rangle_{\rm max, halo} = 16.32\,\mu$G. We note that the value of $X=0.322$ for Coma (Colafrancesco et~al. 2014) is approximately twice the vaue of $\delta p/p = 0.15$ value determined by Fusco-Femiano et~al. (2013) for Coma. These values are consistent given that $p_{\rm e}/p \approx 0.5$. 

Our equipartition value for the Coma cluster is consistent with that of Thierbach et~al. (2003), who find $B_{\rm min, eq}=0.68\,\mu$G (with $k=1$), allowing for varying cosmologies. Although it has been argued that this value is an underestimate, due to the choice of $k$: Beck \& Krause (2005) suggest that this field strength could be as high as 4\,$\mu$G, assuming $n_{\rm p}/n_{\rm e} = 1000$. Magnetic field strength measurements for Coma have also been made using the Faraday rotation of polarized emission from its galactic population (Bonafede et~al. 2013). Unlike minimum energy equipartition measurements, Faraday rotation, $\phi$, provides a direct measure of the magnetic field strength along the line of sight (l.o.s.) such that $\phi \propto \int_{\rm l.o.s.}{n_{\rm e} B_{\rm ||} {\rm d}\ell}$. For the Coma cluster, Faraday rotation measurements indicate that the average l.o.s. magnetic field strength within $r_{\rm c}$ is $B_{\rm ||}=4.7\,\mu$G (Bonafede et~al. 2010), consistent with the limits set here using equipartition and the non-thermal pressure fraction. 

We note that the calculations outlined here and in Section~\ref{sec:scaling} assume that the clusters under examination are well described by a $\beta$-model. This assumption creates limitations in the situation where either the cluster gas density or non-thermal halo gas population deviates significantly from spherical symmetry. In this situation the estimates for the non-thermal pressure (and hence the magnetic field strength) are likely to differ. There are two potential causes for such a situation in this case: firstly, this is a merging system and the assumed beta model is may not be a good representation; secondly, a possible unresolved radio relic could bias the radio power high. Given the low radio luminosity of the proposed Triangulum Australis halo relative to the general scaling relations, see Fig.~\ref{fig:scaling}, the second of these scenarios seems unlikely; however, observations at higher resolution with improved sensitivity relative to currently available data are required to examine this possibility in more detail. The former scenario is likely to affect the results presented here, but to what extent is currently unclear. Further development of the methodology used to calculate the non-thermal fraction will be necessary to assess the impact.

\section{Conclusions}\label{conclusion}

We have used new observations with the KAT-7 telescope to make the first detection of a diffuse radio halo in the Triangulum Australis cluster. By combining these new radio data with complementary data in the X-ray and SZ regimes, we have demonstrated that this cluster is consistent with the established scaling relations for clusters hosting haloes. In addition we have:

\begin{itemize}
\item Used a combination of X-ray and SZ data to determine the ratio of non-thermal to thermal pressure within the cluster, which we determine to be $X=0.658\pm0.054$.
\item From this ratio of pressures we were able to determine an upper limit on the average magnetic field strength within $R_{500}$, $\langle B\rangle_{\rm max, 500} = 19.46\,\mu$G, and hence within the halo region, $\langle B\rangle_{\rm max, halo}=33.08\,\mu$G.
\item We have compared these values with the lower limit equipartition value determined from the radio power, under stated assumptions, which we determine to be $B_{\rm min} = 0.77(1+k)^{2/7}\,\mu$G. Hence providing both lower and upper limits on the possible field strengths within the cluster halo region.
\item We use the well-studied Coma radio halo to contextualize these results and demonstrate that the range of values we calculate for the allowable magnetic field strengths are consistent with measurements made using alternative methods.
\end{itemize}

\vspace{-1.0em}

\section*{Acknowledgments}

We thank the staff of the Karoo Observatory for their
invaluable assistance in the commissioning and operation of the
KAT-7 telescope. The KAT-7 is supported by SKA South Africa 
and the National Science Foundation of South Africa. AMS gratefully acknowledges support
from the European Research Council under grant ERC-2012-StG-307215 LODESTONE.

\end{document}